\documentstyle[sprocl]{article}
\begin{document}
\newcommand{\bi}{\bibitem}
\newcommand{\be}{\begin{equation}}
\newcommand{\ee}{\end{equation}}
\newcommand{\bt}{\begin{table}}
\newcommand{\et}{\end{table}}
\newcommand{\btab}{\begin{tabular}}
\newcommand{\etab}{\end{tabular}}
\def\eg{{\it e.g.}}
\def\ra{\rightarrow}
\def\dek#1{\times10^{#1}}

\def \ea{{\it et al.}}
\def \appb#1#2#3{Acta Phys. Pol. B {\bf#1}, #2 (19#3)}

\font\eightrm=cmr8

\input{psfig}

\bibliographystyle{unsrt} 

\arraycolsep1.5pt

\def\Journal#1#2#3#4{{#1} {\bf #2}, #3 (#4)}

\def\NCA{\em Nuovo Cimento}
\def\NIM{\em Nucl. Instrum. Methods}
\def\NIMA{{\em Nucl. Instrum. Methods} A}
\def\NPB{{\em Nucl. Phys.} B}
\def\PLB{{\em Phys. Lett.}  B}
\def\PRL{\em Phys. Rev. Lett.}
\def\PRD{{\em Phys. Rev.} D}
\def\ZPC{{\em Z. Phys.} C}

\def\st{\scriptstyle}
\def\sst{\scriptscriptstyle}
\def\mco{\multicolumn}
\def\epp{\epsilon^{\prime}}
\def\vep{\varepsilon}
\def\ra{\rightarrow}
\def\mee{M_{ee}}
\def\ppg{\pi^+\pi^-\gamma}
\def\pmm{\pi^+\mu^+\mu^-}
\def\kpmm{K^+\rightarrow\pi^+\mu^+\mu^-}
\def\kpme{K^+\rightarrow\pi^+\mu^+ e^-}
\def\pee{\pi^+e^+e^-}
\def\kpee{K^+\rightarrow\pi^+e^+e^-}
\def\vp{{\bf p}}
\def\ko{K^0}
\def\kb{\bar{K^0}}
\def\al{\alpha}
\def\ab{\bar{\alpha}}
\def\be{\begin{equation}}
\def\ee{\end{equation}}


\title{ $\kpmm$ IN E865 AT BNL}
\author{ JULIA A. THOMPSON}
\author{ R.A. APPEL, D.N. BROWN, N. CHEUNG,
C. FELDER, M. GACH}
\author{ D.E. KRAUS, P. LICHARD,  A. SHER}
\address{Dept. of Physics and Astronomy, University of Pittsburgh\\
Pittsburgh, PA., 15260, USA\\e-mail:jth@vms.cis.pitt.edu}
\author{D.M. LAZARUS, H. MA, P. REHAK}
\address{Brookhaven National Laboratory, Upton, NY, 11973, USA}
\author{G.S. ATOYAN, V.V. ISSAKOV, A.A. POBLAGUEV}
\address{Institute for Nuclear Physics, Moscow, Russia}
\author{H. KASPAR}
\address{Paul Scherrer Institute, CH-5232 Villigen, Switzerland}
\author{B. BASSALLECK, S. EILERTS, H. FISCHER, J. LOWE}
\address{University of New Mexico, Albuquerque, New Mexico, 87131, USA}
\author{S. PISLAK, P. TRUOEL, University of Zurich and Yale University}
\author{D. BERGMAN, S. DHAWAN, H. DO, J. LOZANO, W. MAJID, M. ZELLER}
\address{Yale University, New Haven, Connecticut, 06520-8121, USA}

\maketitle\abstracts{
Preliminary values for the $\kpmm$ branching ratio and form factor are
reported, based on 400
 events, a factor of 2 more in total events and 
 100 times  the present world sample of fully reconstructed events.
   The results are consistent with previous results on the $\pee$ mode.
However, the relatively large  slope of the form factor in $q^2$,
$\lambda = 0.182 \pm 0.01 \pm 0.007 $,  required to fit the
$\pee$data and to give consistency between the
$\pee$ and $\pmm$ branching ratios, is larger than expected
in simple models of the decays.  The $\kpmm$ branching
ratio we find, $(9.23 \pm 0.6_{stat}\pm 0.58_{syst.})\times 10^{-8}$
is the most precise measurement of this mode and is $\approx
3.2 \sigma$  larger than the previous measurement.
These $\pi l l $ results are inconsistent with
$O(p^4)$ Chiral Perturbation Theory but  compatible with $O(p^6)$.
Systematic studies for both modes are still in progress. 
}
\section*{Introduction and Theoretical Background}
 The primary  interest of E865 is the forbidden decay $\kpme$.
Results from two independent analyses
 \cite{db,sp} have yielded a limit of
 2 $\times~10^{-10}$  for a preliminary 1995 data set,
   data from 1996 and 1998 has a 
  statistical reach 
 of order $10^{-11}$.  However, an important by-product of this experiment
 has been that E865 has significantly increased world sample sizes
 for several  other $K^+$ decay modes
with 3 charged particles in the final state.
 (Final states containing a $\pi^{0}$ are included, since the  $\pi^{0}$ is
detected through
$\pi^{0} \rightarrow \gamma e^{+} e ^{-}$). A detailed understanding
of the K decays, among them $\pee$ and $\pmm$, tests models of the
weak interaction and low energy QCD.

 The focus of this paper is a new measurement of the branching fraction
 for $\kpmm$.
 The previous best measurement, from E787 \cite{kmm787},
 based on a small number
of fully reconstructed events and about 200 partially reconstructed
events, is
 $(5.0 \pm 1.0)\times 10^{-8}$; they did not make a form factor
  measurement, because of  limited
  acceptance and incomplete event information.
  With specific assumptions about the form of the interaction
  (typically, vector, with a linear q$^2$ dependence, as in Ke3),
  the expected  $\pmm / \pee$ ratio can be
  calculated from the E865 results, and from theory, and 
   compared with the experimental observation.
  For the $\pee$  we use the most precise results,
   a branching fraction of $(2.82\pm 0.04\pm 0.15)10^{-7}$
   and $\lambda$ of $0.182 \pm 0.01 \pm 0.007$, 
   based on our 10000 events  \cite{se}.

Short distance contributions \cite{short} to $K \rightarrow
\pi \it{l} \it{l}$ are only of order $10^{-9}$.
Long distance contributions \cite{bs,pl,eils,vain}
come close to the observed $\pee$ rate and give a $\pmm$ to $\pee$
ratio of about 0.22-0.24.
Vector and a1 meson dominance \cite{pl}
is an example of a simple model with a
parameter free prediction for the branching fractions but  small
form factor dependence on $q^2$, similar to  Ke3.
 Chiral Perturbation theory parameters
allow  a range of predictions for the form factor. In  $O(p^4)$
\cite{epr,gabbiani},
the form factor and its $q^2$ dependence
 are tightly correlated with the branching ratio.
Using an  $O(p^6)$  calculation of an explicit "pion loop term" \cite{dam}
 with a polynomial of the expected Chiral Perturbation
 behavior at $O(p^6)$
  gives additional parameters and   flexibility.

  The E787
    $\pmm / \pee$ ratio, $ 0.18 \pm 0.04$, is
 about 1.5 $\sigma$ below predictions.

\section*{Experimental Apparatus}
E865 at BNL \cite{hmavan} is a magnetic spectrometer
 illuminated by an intense
unseparated 6 GeV/c beam, of
$\approx 10^{8} K^{+}$ and
$\approx 2 \times 10^{9} \pi^{+}$ per 1.6 sec AGS pulse.
 Momentum measured in the spectrometer is compared  with
 energy deposit in the 600 module 15 r.l. deep "Shashlik" calorimeter.
   Electron
and positron identification is done by two  threshold
Cerenkov counters  with $H_{2}$ on the left  (primarily 
negative particles) and $CH_{4}$ on the right  (primarily
positive particles). A  24 plane  proportional tube - iron plate
range stack identifies muons.
  The  $\pee$ data were taken parasitically in 1995 and 1996, and
  $\pmm$ data in  a   1997 reduced intensity run. 

\section*{Event Selection and Analysis}
  The $\pmm$ events are normalized
to the  $\pi^+\pi^+\pi^-$ final state, with similar kinematics. 
The trigger for all  modes \cite{hmavan} requires
three particles in a kinematically plausible configuration
 in hodoscope counters and the calorimeter.
The analysis required: a good reconstructed vertex; reasonable vector
momentum; and  two electrons or two muons,
  one negatively charged
 on the left and one positively charged on the right.
For the $\pee$ events,  Cerenkov counter light is required both in the
trigger and in the analysis; and for the $\pmm$ events,  muon chamber
signals  were. 
Cuts in track chisquare  help eliminate
the primary background for the $\pmm$ events, 
secondary decays ($\pi \ra \mu \nu$)
 from $K^{+} \ra \pi^+ \pi^+ \pi^-$ .

\begin{figure}[htb]
\centerline{
\psfig{figure=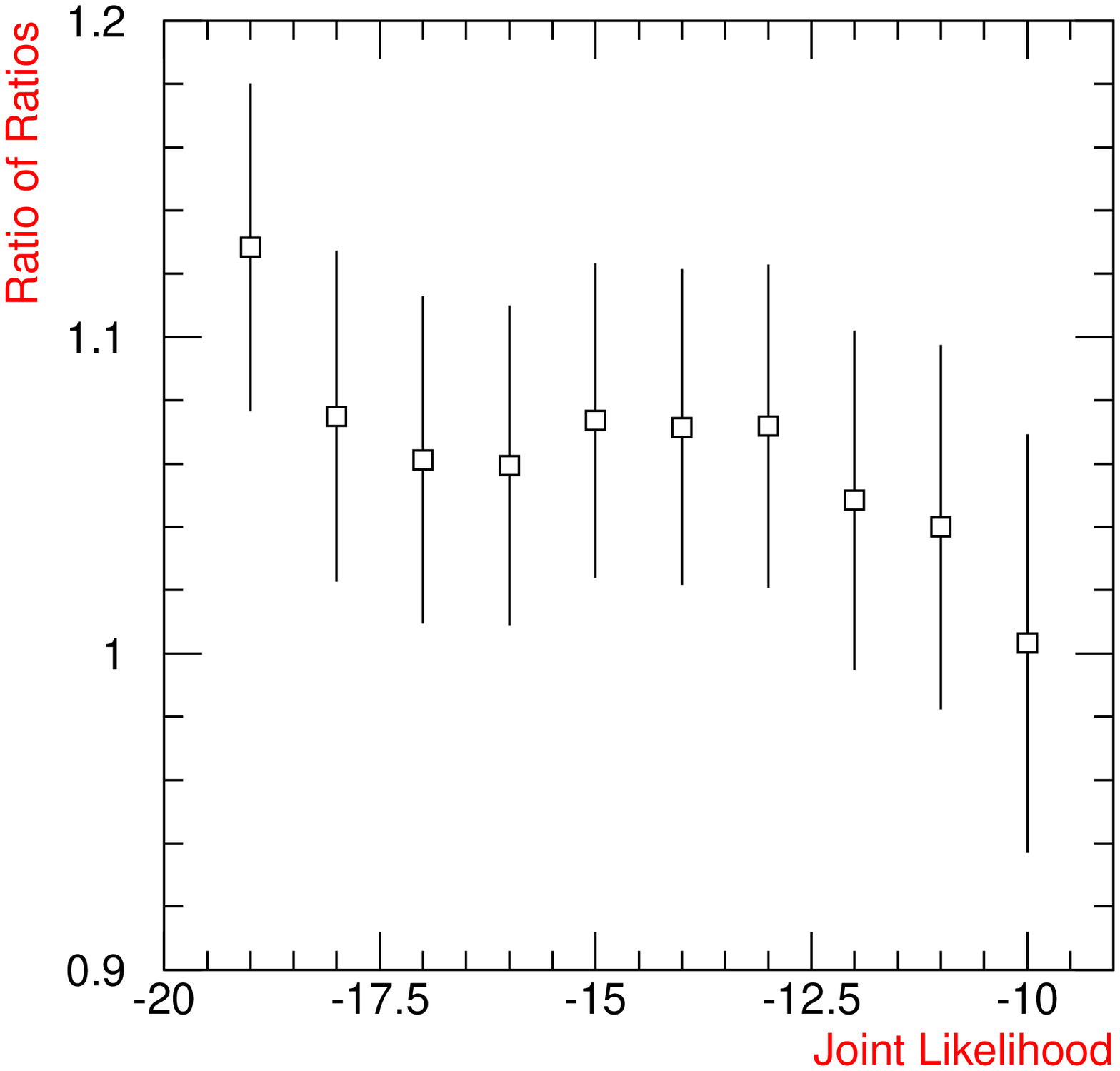,height=2.3in,width=2.3in}
\psfig{figure=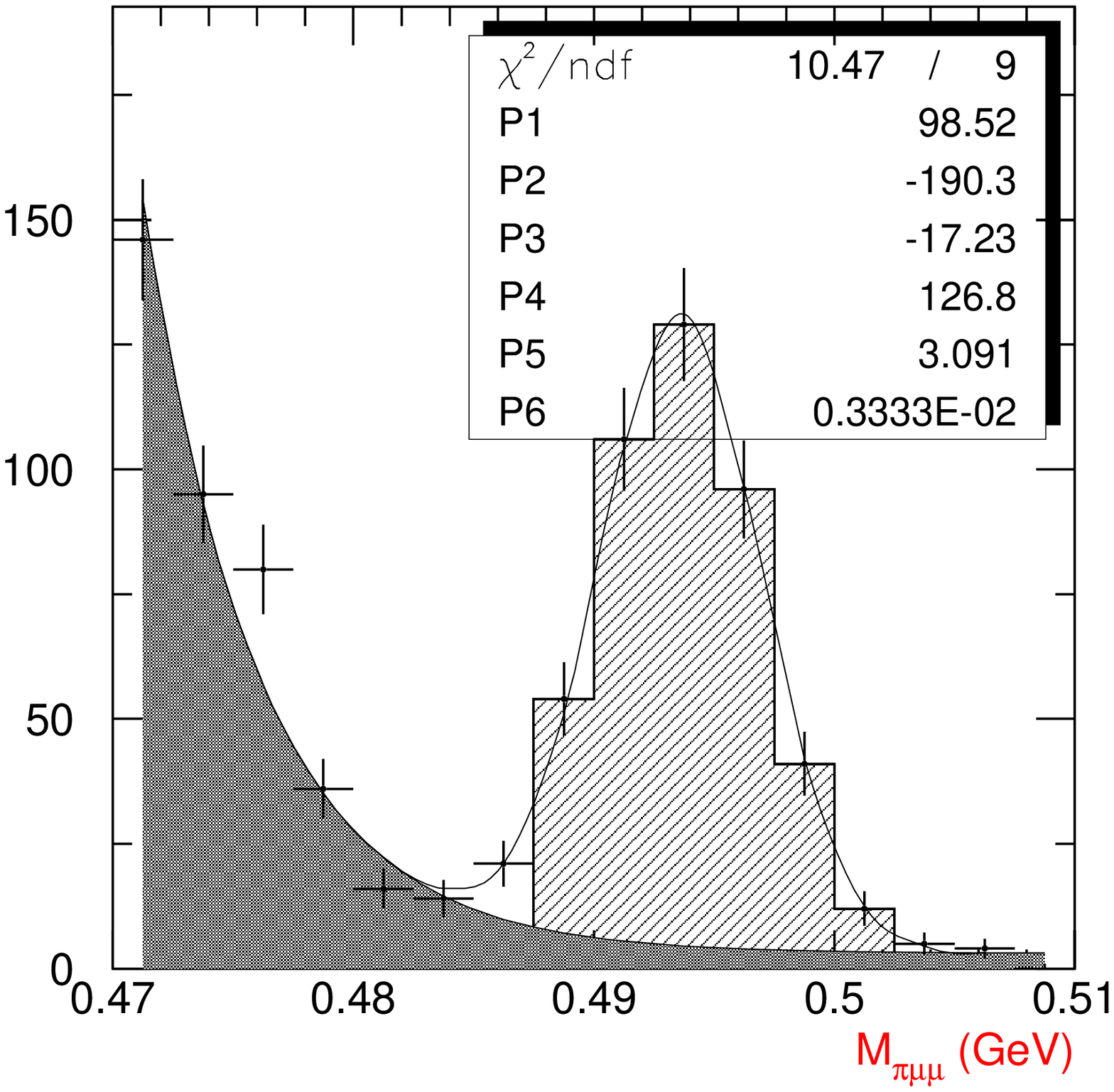,height=2.3in,width=2.3in}
}
\caption[]{ 
Left: R as a function of the joint likelihood cut,
and track $\chi^2$ cut.
Right:$\pmm$ invariant mass distribution for
     joint likelihood cut of -13. 
}
\label{fig:stabil}
\end{figure}

The information described
qualitatively above was combined quantitatively into an
"event likelihood".
The stability of the  branching fraction as a function of
 event likelihood
is shown in the  plot on the left in Figure \ref{fig:stabil},
where the quantity R (proportional to the
branching fraction) is plotted against the event likelihood cut.
 R is defined as:  $r_{\mu}/r_{3\pi}$ where
$r_{\mu~ (or ~ 3 \pi)} = Ndata_{\mu (or~ 3 \pi)}/Nmc_{mu (or~ 3\pi)}$
and Ndata and Nmc are respectively the accepted data (simulated) events.
While R is relatively stable, the number of signal events (not shown
because of space constraints) drops from 700  to 200
 as the event likelihood moves from -19 to -10.  This drop
 in signal is accompanied by a drop in background to signal
 from $\approx 40 \%$   to $\approx 2\%$ , reflecting the
 large admixture of background at the large negative
 value of the event likehood, and loss of signal as the
 event likelihood approaches -10.  For our final result,
 we use an  event likelihood cut of -13,
 which gives  $\approx$ 400 signal events. 

The effective mass of the $\pmm$ final state is shown in
the right hand side of Figure \ref{fig:stabil}.
Background from  $K^+ \ra \pi^+\pi^+\pi^-$, with two
pions decaying to muons, is shown by the dark-shaded
curve at low effective masses.  
The systematic error is estimated as $\approx 7\%$,
dominated by $\approx 3\%$ from selection criteria
and background subtraction, and $\approx 4\%$ from
 normalization uncertainties.

%



\section*{Results and Discussion}

\begin{figure}[htb]
\centerline{
\psfig{figure=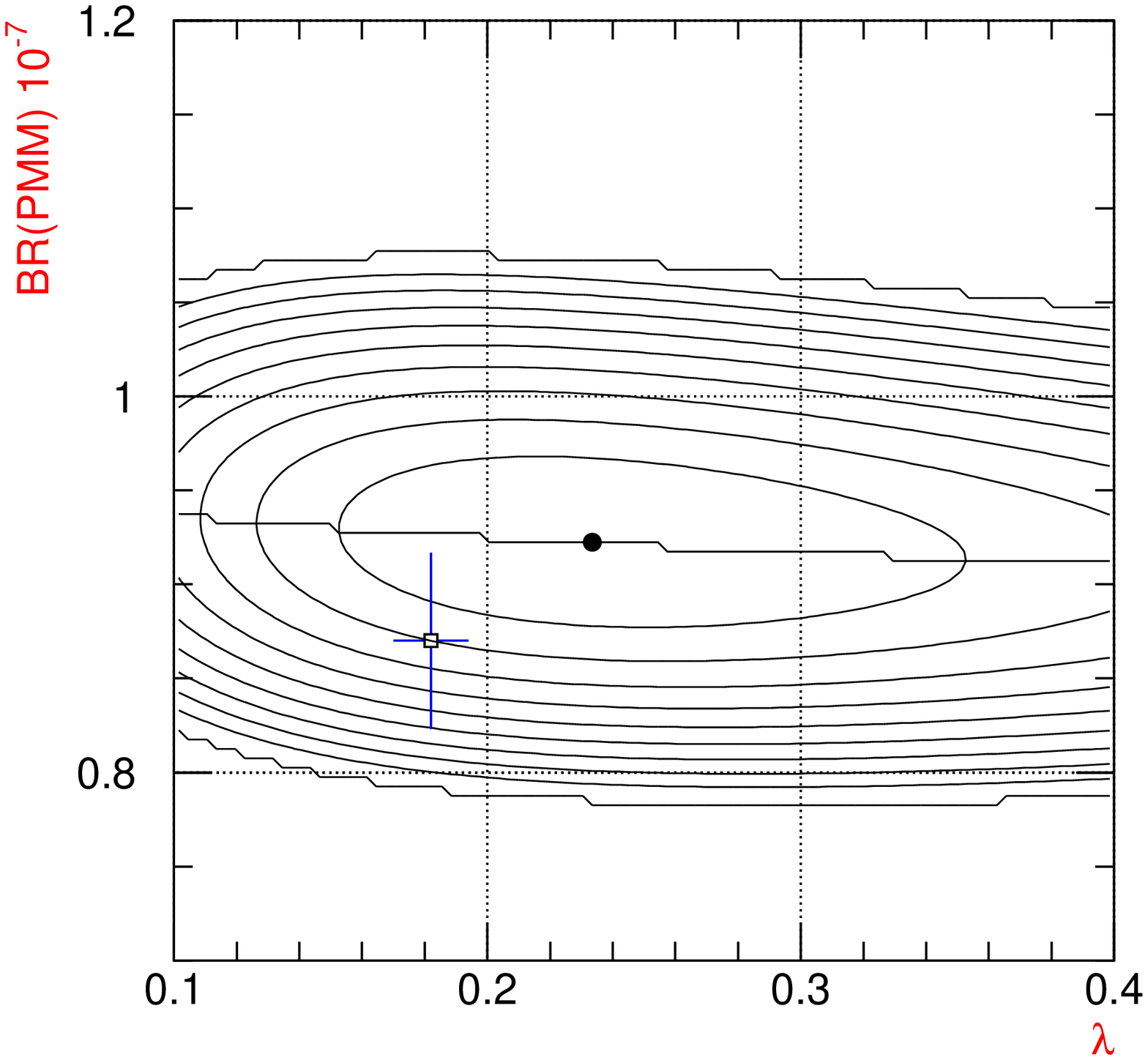,height=2.3in,width=2.3in}
\psfig{figure=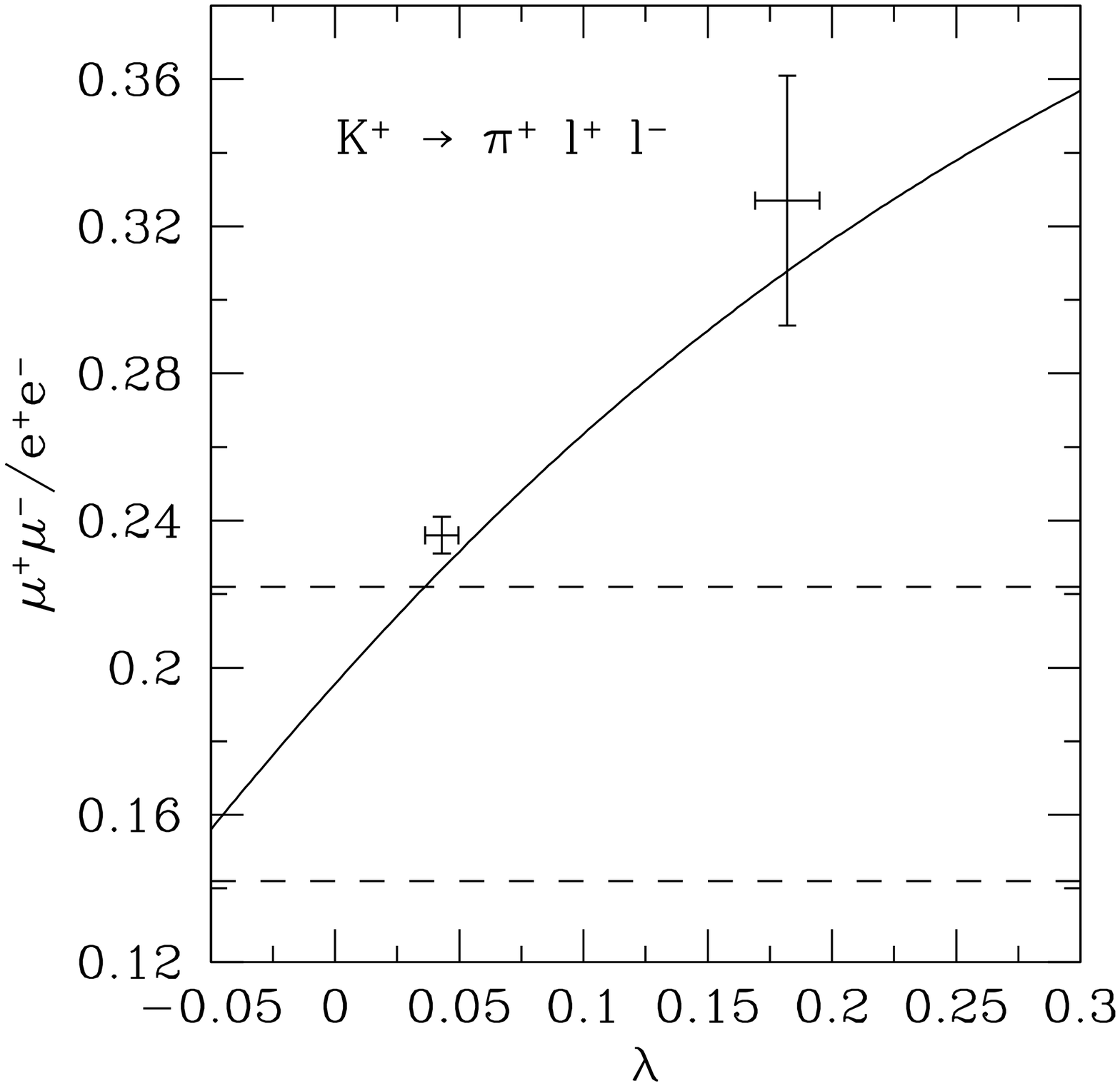,height=2.1in,width=2.1in}
}
\caption{
Left: $\chi^{2}$ contour of Br vs $\lambda$.
Each contour is one unit of $\chi^{2}$. 
The solid point is  the  $\chi^{2}$ minimum, 
and the open point with error bars
is the prediction from the $\pee$ measurement.
Right:Relative branching ratios of the $\pmm$ and $\pee$ decay
modes vs $\lambda$, assuming a linear dependence of the
form factor on $q^2$. The previous result is shown as a band,
while the point at low $\lambda$ is from Lichard's  meson dominance model,
as representative
of simple long distance models.
The point at high $\lambda$ is the result presented in this paper, using
the $\lambda$ and the $\pee$ branching fraction from
the thesis of Scott Eilerts, the most precise determination
of these parameters.
}
\label{fig:lambda}
\end{figure}

The $\mu\mu$ effective mass distribution ($q^2$) and
the
 $\cos\theta_{\pi\mu^+}$ distribution (where $\theta_{\pi\mu^+}$ is the
angle between the $\pi^+$ and $\mu^+$ in the $\mu\mu$ center of mass frame)
are not shown but are consistent with a vector interaction
in the decay, with a form factor $\approx 0.2$, as seen in the $\pee$ events
\cite{hmavan,se}.
Fit contours for the  branching fraction and $\lambda$
(the form factor $q^2$ dependence) are shown in Figure \ref{fig:lambda},
assuming a linear $q^2$ dependence as in Ke3.
The   $\pmm$
 branching fraction and $\lambda$ are larger than expected
in a simple meson dominance model but agree with
expectations from $\pee$.
Our present understanding is that our $\pi l l $ data (branching ratios, and
q$^2$ dependence,
taken together, Ref. \cite{hmavan,se} and from this paper)
 are inconsistent 
with Chiral Perturbation theory at $O(p^4)$ 
\cite {epr,gabbiani},
but, due to the additional parameters available,
are consistent with 
an O($p^6$) calculation \cite{dam}.
Detailed
systematic studies and comparison with Chiral Perturbation
theory are in progress.

\section*{Acknowledgments}
We  thank our home country research institutes 
for  financial support.  The experiment
would not have been possible without the valiant and sustained
efforts of the AGS machine operators and technical support crew.
The Pitt group owes a large debt of gratitude to high school
teacher Ivan Ober and  
many students who helped in  design, construction and
commissioning of the Cerenkov counters, especially Elizabeth
Battiste, Rebecca Chapman, Amy Freedman, Tuan Lu, Cindy Miller,
Melinda Nickelson, Tim Stever, Paula Pomianowski, and Craig Valine.


\section*{References}
\begin{thebibliography}{99}
\bibitem{kmm787}S. Adler {\it{et al.}}, \Journal{\PRL}{79}{4756-4759}{1997}.

\bibitem{e777}C. Alliegro {\it et al.},\Journal{\PRL}{68}{278}{1992}.

\bibitem{db}D.R. Bergman,PhD Thesis,
     Yale University, December, 1997.

\bibitem{bs} L. Bergstrom, P. Singer, \Journal {\PRL}{55}{2633}{1985} and
    \Journal{\PRD}{43}{1518}{1991}.

\bibitem{e777pme}C. Compagnari,  PhD Thesis,
    Yale University (1988).
\bibitem{dam}G. D'Ambrosio, G. Ecker, G. Isidori, J. Pertoles, JHEP {\bf 8},
   4, 1998.
\bibitem{gabbiani}J.Donoghue, F. Gabbiani, \Journal{\PRD}{51}{2187}{1995}.

\bibitem{epr}G. Ecker, A. Pich, E. de Raphael, \Journal{\NPB}{291}{692}{1987}.

\bibitem{eils}G. Eilam and M.D. Scadron, \Journal{\PRD}{43}{1568}{1985}.

\bibitem{se} S.W. Eilerts,
 PhD Thesis, University of New Mexico, 1998.


\bibitem{short} for example, M.K. Gaillard and B.W. Lee,
    \Journal{PRD}{10}{897}{1974};
     D.V. Nanopoulous and G.G. Ross, \Journal {\PLB}{56}{279}{1975};
     P. Agrawal, {\it{et al.}},\Journal{\PRD}{45}{2383}{1992}.

\bibitem{pl}P. Lichard, \Journal{\PRD}{55}{5385}{1997} and hep-ph/9903216.

\bibitem{hmavan}H.Ma et al., {\it Proceedings of the XXIX International
Conference on High Energy Physics, Vancouver, 1998} (World Scientific,
Singapore, in press).

 \bibitem{sp} Stefan Pislak,Inaugural-Dissertation,
 University of Zurich, 1997.

\bibitem{vain} A.I. Vainshtein, V.I. Zakharov, L.B. Okun, and M.A. Shifman,
 Yad. Fiz. {\bf 24},820 (1976). [Sov. J. Nucl. Phys. {\bf 24},427 (1976)].

\end{thebibliography}
\end{document}